\documentclass[fleqn,twoside]{article}
\usepackage{espcrc2}

\pdfoutput=1
\usepackage{graphicx}
\usepackage[figuresright]{rotating}

\usepackage{subfigure} 
\usepackage{amsmath}   
\usepackage{units}
\usepackage{accents}

\def\etal{\textit{et al.}}

\title{Studies of the UHECR propagation in the Galactic Magnetic Field}

\author{S. Vorobiov\address[UNG]{Laboratory for astroparticle physics, 
        University of Nova Gorica, Slovenia},
        M. Hussain\addressmark[UNG], and
        D. Veberi\v{c}\addressmark[UNG]\address{J.\ Stefan Institute, Ljubljana,
        Slovenia}}
       
\begin{document}

\begin{abstract}
We present the results of simulations of the ultra-high energy cosmic ray 
(UHECR) propagation in the Galactic magnetic field (GMF).
Different assumptions on the large-scale GMF structure and/or primary 
particle lead to distinctly different deflection patterns 
of the highest energy cosmic rays (CR). The GMF modifies 
the exposure of an UHECR experiment to the extragalactic sky.
We estimated these effects for the Pierre Auger experiment.
Further forward-tracking studies under plausible UHECR sources scenarios 
will allow for direct comparison with the observed correlation 
between the nearby active galactic nuclei (AGN) and the highest energy Auger events.
\end{abstract}

\maketitle

\section{Introduction}

Magnetic fields in the Milky Way are investigated
via Faraday rotation of the polarized light from pulsars 
and extragalactic (EG) sources, and through measurements 
of the Galactic synchrotron radiation~\cite{beck,widrow}.
The observations reveal the presence of a large-scale (LS) 
Galactic magnetic field (GMF), 
that in the first approximation follows the spiral arm structure. 
The LS field extends above and below 
the Galactic disk on a kiloparsec scale and forms a kind of halo.
A random GMF, of the strength similar 
to that of the regular component, but with the coherence length
of only $\sim \unit[50]{pc}$, has also been observed~\cite{beck}. 
The average strength of the total magnetic field 
near the Sun is about $6\,\mu{\rm G}$. However, 
details of the field distribution are poorly 
known~\cite{han,brown,men}.

The Pierre Auger Observatory~\cite{auger} provides a new and independent way 
of studying cosmic magnetic fields, by collecting CR events 
above $\unit[10]{EeV} \equiv \unit[10^{19}]{eV}$ with
unprecedented statistics and data quality. Recently, Auger 
observed a significant correlation (over angular scales $\leq 6^\circ$)
of the arrival directions of CR above $\simeq \unit[60]{EeV}$ 
with the locations of nearby AGN~\cite{augeragncorrelationSci,augeragncorrelationAPh}, 
and a strong steepening of the CR flux above $\unit[40]{EeV}$~\cite{augerspectrumPRL}.
Both observations are consistent with the standard 
UHECR astrophysical acceleration scenarios 
and thus represent an important step 
towards the ``charged particle astronomy''. 

The elaboration of relevant analysis methods 
for the UHECR astronomy requires detailed investigation 
of the cosmic ray propagation in the GMF.
We present in this paper results of such studies,
performed in the light of the observed AGN correlation.
We used the standard method of CR backtracking. 
A large number of CR events has been simulated 
using parameters described below, and propagated 
under three distinctive large-scale GMF models.
We present the resulting magnetic deflections  
and modification of the extragalactic exposure 
due to the LS GMF.

\section{Implementation of CR backtracking}
\label{s:Backtracking}

To obtain the UHECR trajectories,
we implemented the integration of the equations of motion of an 
ultrarelativistic particle in the magnetic field 
using the Runge-Kutta $5^\text{th}$
order scheme with the adaptive step size control 
~\cite{numerical}. The particles have been followed till the galactocentric 
distance of $\unit[20]{kpc}$ (the Galaxy ``border''), beyond which 
the GMF strength is supposed to be negligibly small.
An accuracy level (the accepted truncation error) of $10^{-6}$ 
has been adopted, for which the changes in the backtracked directions 
become negligible with respect to the chosen step in angular separation $d_{\text{max}}$ 
from the selected astrophysical objects (see the section~\ref{s:ScanParameters}).
To avoid the ``numerical dissipation'' of 
the CR energy~\cite{armengaud}, 
we preserved the absolute value of the 
CR velocity vector during propagation.

\begin{figure*}[!ht]
\centerline{\subfigure[Bisymmetric even parity spiral field]
{\includegraphics[width=0.45\textwidth]{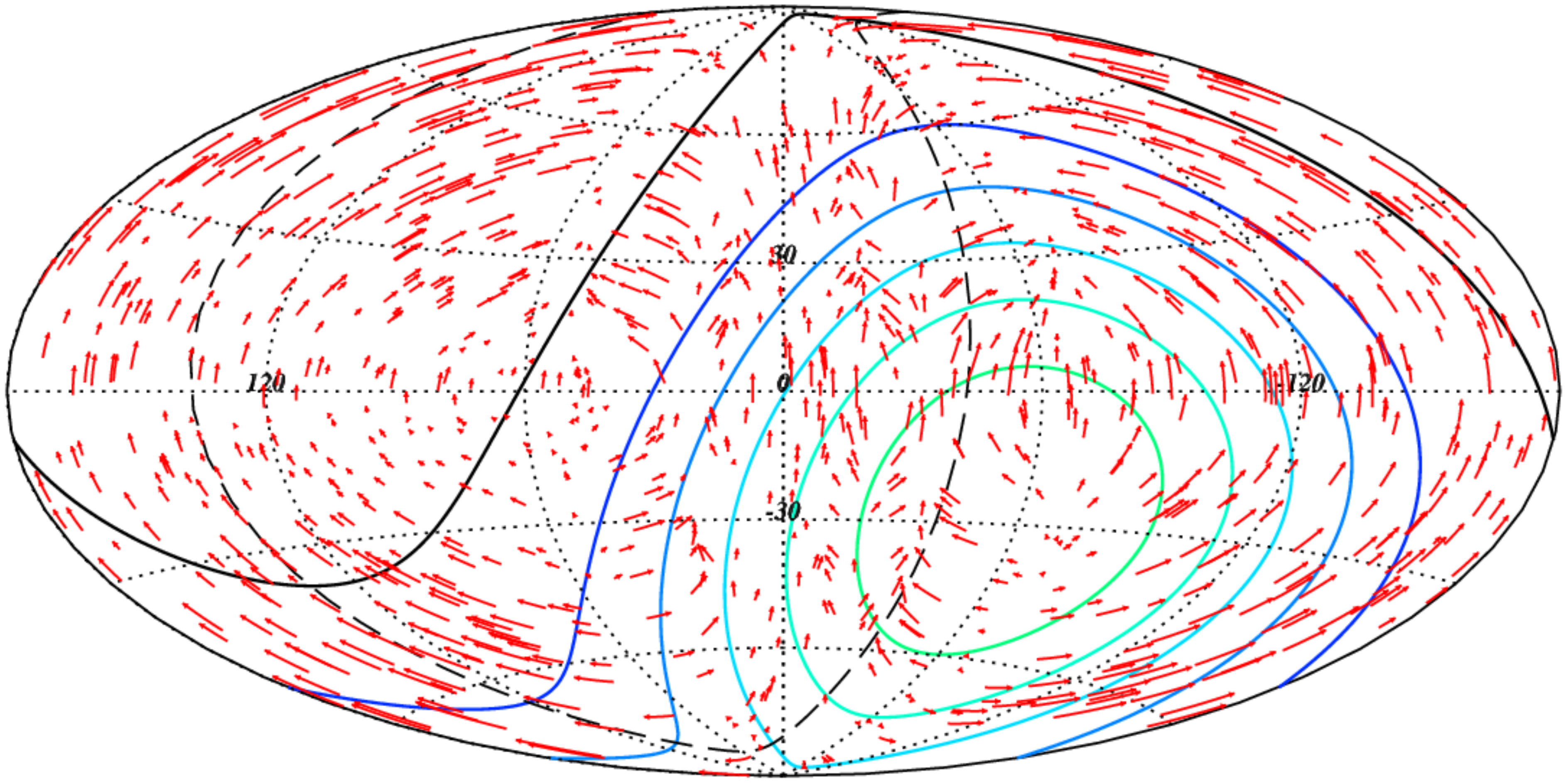}
\label{fig:hmrbsss_defldir}}
\hfil
\subfigure[Axisymmetric odd parity spiral field]
{\includegraphics[width=0.45\textwidth]{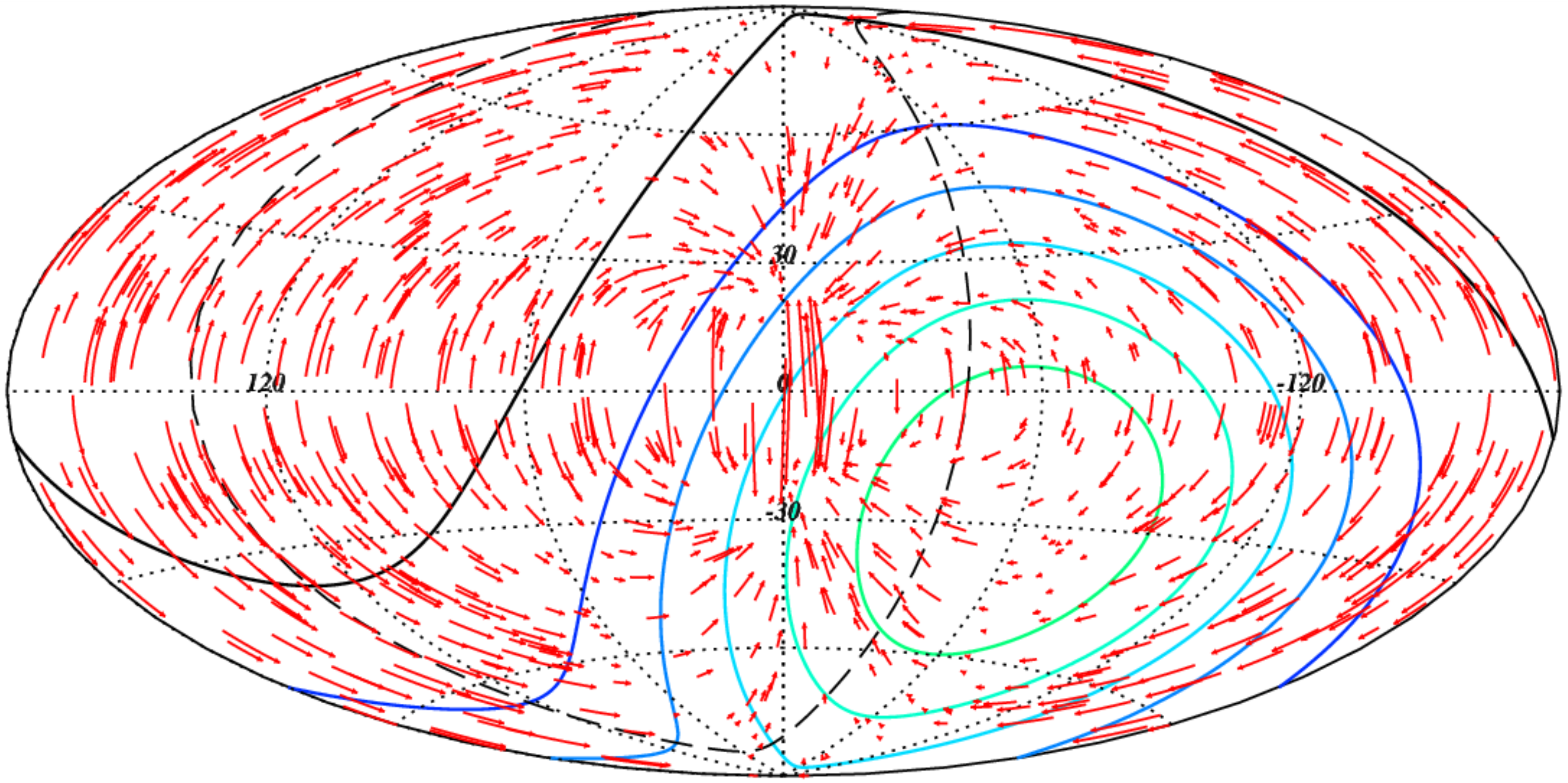}
\label{fig:hmrassa_defldir}}}
\caption{Cosmic ray deflection patterns for \emph{protons} 
under Models~I and~II. Arcs of great circles join the simulated 
arrival directions on Earth, and the backtracked ones at the Galaxy border 
(arrow heads). Every $1000^{\text{th}}$ event is shown.  
The dashed line denotes the supergalactic plane (SGP). The solid lines 
show the equal integrated Auger SD exposure sky regions within the detector 
field of view for $\theta_\text{z} \le 60^\circ$.}
\label{fig:hmrmodels_defldir}
\end{figure*}

\section{Parameters of simulated CR events}
\label{s:UHECRParameters}

\textbf{Energies} of the simulated events were bounded 
between \unit[40]{EeV} and \unit[150]{EeV}, and distributed
according to the Auger measurement~\cite{augerspectrumPRL}, 
as a power law $E^{-4.2}$. 

Two options for the \textbf{angular distribution} of the events have been 
considered: 
1) Uniform over the whole sky (10$^6$ events), and 
2) According to the exposure of 
the Auger Surface Detector (SD) for zenith angles 
$\theta_\text{z} \le 60^\circ$~\cite{sommers,AugerSDExposure} (10$^5$ events).

Four pure \textbf{mass compositions} have been considered:
protons, carbon, silicon, and iron nuclei. 

\section{Considered large-scale GMF models}
\label{s:GMFModels}

We have chosen amongst many available large-scale GMF models \textit{three}
typical ones with distinct qualitative differences. 
Two models are the spiral disk field models of~\textit{bisymmetric even
parity} (hereafter \textbf{Model~I}), and \textit{axisymmetric odd parity} 
(\textbf{Model~II}) from the paper~\cite{harari}
by Harari, Mollerach and Roulet (HMR).

\textbf{Model~III} is a modification of the model \cite{prouza}
by Prouza and \v{S}m\'ida (PS), that has been proposed by 
Kachelrie\ss~\etal~\cite{kachelriess}. In addition to the 
\textit{bisymmetric even parity} spiral field, of the structure
similar to the one in Model~I, it features 
two additional large-scale halo GMF components: 
\textit{toroidal} azimuthal field above and below the Galactic disk, 
and \textit{poloidal} (dipole) field.

\section{Magnetic deflections of cosmic rays}
\label{s:MagneticDeflections}

The cosmic ray deflection patterns obtained using the three 
selected GMF models are very different, and have trends, typical
for each model. Fig.~\ref{fig:hmrmodels_defldir} shows
deflections of primary \emph{protons} for Models I and II.  
In the bisymmetric even parity field, 
CR arrival directions at the Galactic border point 
in general to higher Galactic latitudes than observed on Earth.  
In the odd parity field, the
backtracked arrival directions in each Galactic hemisphere are
shifted to the poles, and no cosmic rays are coming 
from the directions close to the Galactic plane. The presence 
of dipole and toroidal fields in Model III makes 
the deflection pattern even more complex~\cite{vorobiov}.

The deflection values for the three GMF models, and \textit{the Auger SD exposure} 
are summarized in the table~\ref{t:DeflectionAndLog10Pmin}, by
means of percentiles at 50\% (median), and 95.45\% of
the c.d.f. The median deflection values scale well with the
atomic number $Z$ of primary nuclei. They
are larger for the HMR models than in Model III, due to 
the more important halo field extension.

The simulated events allow to roughly estimate the corresponding
deflections expected from turbulent magnetic fields~\cite{vorobiov}.
Compared to the deflections from the large-scale field,
the former ones have to be much smaller, which is confirmed 
by simulations including turbulent field~\cite{tinyakovrandom}, 
and can be neglected at the first approach.

\section{GMF effects on the EG exposure}
\label{s:EGExposureEffects}

\begin{figure}[!ht]
\centering
\includegraphics[width=0.4\textwidth]
{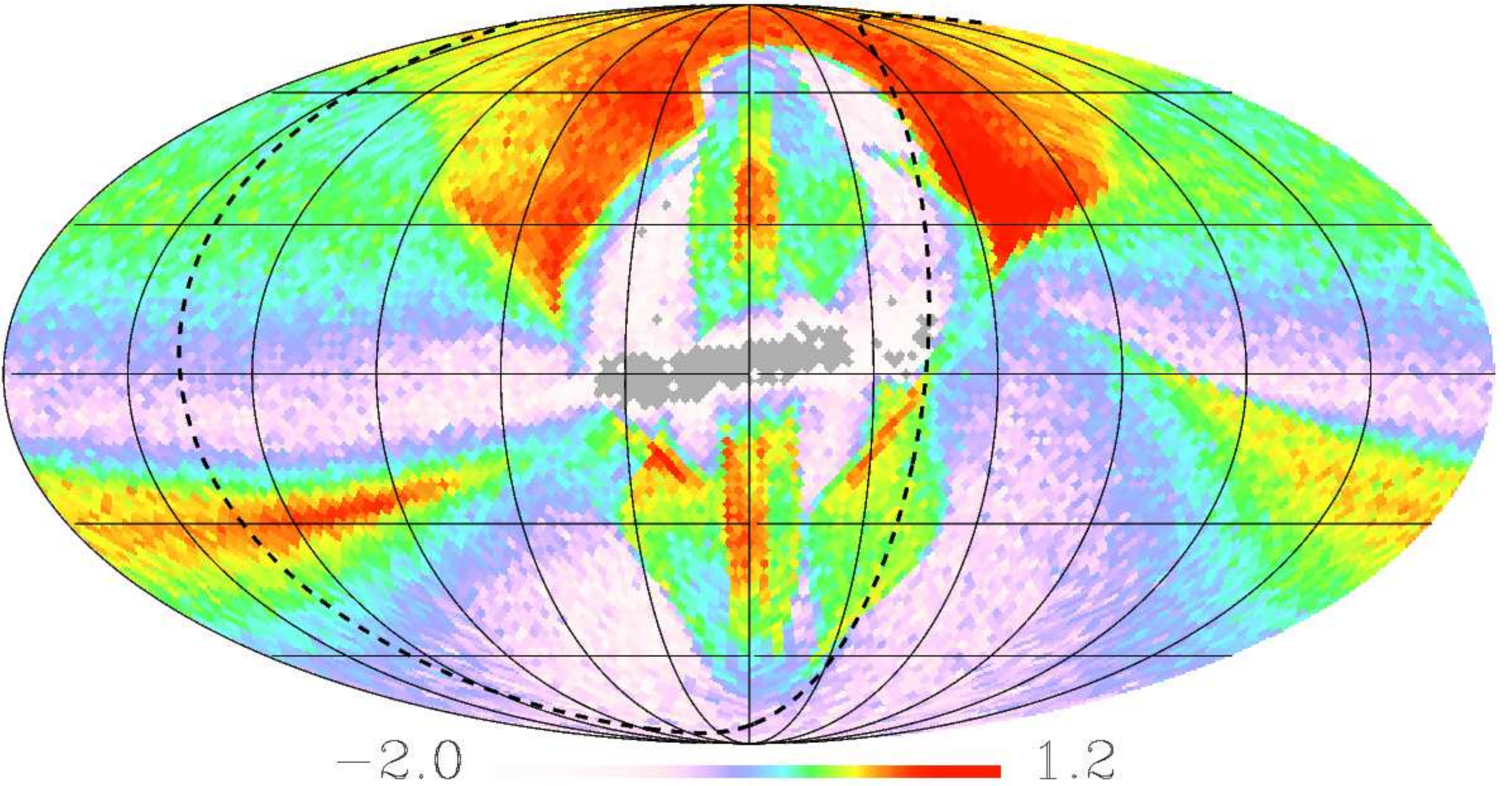}
\includegraphics[width=0.4\textwidth]
{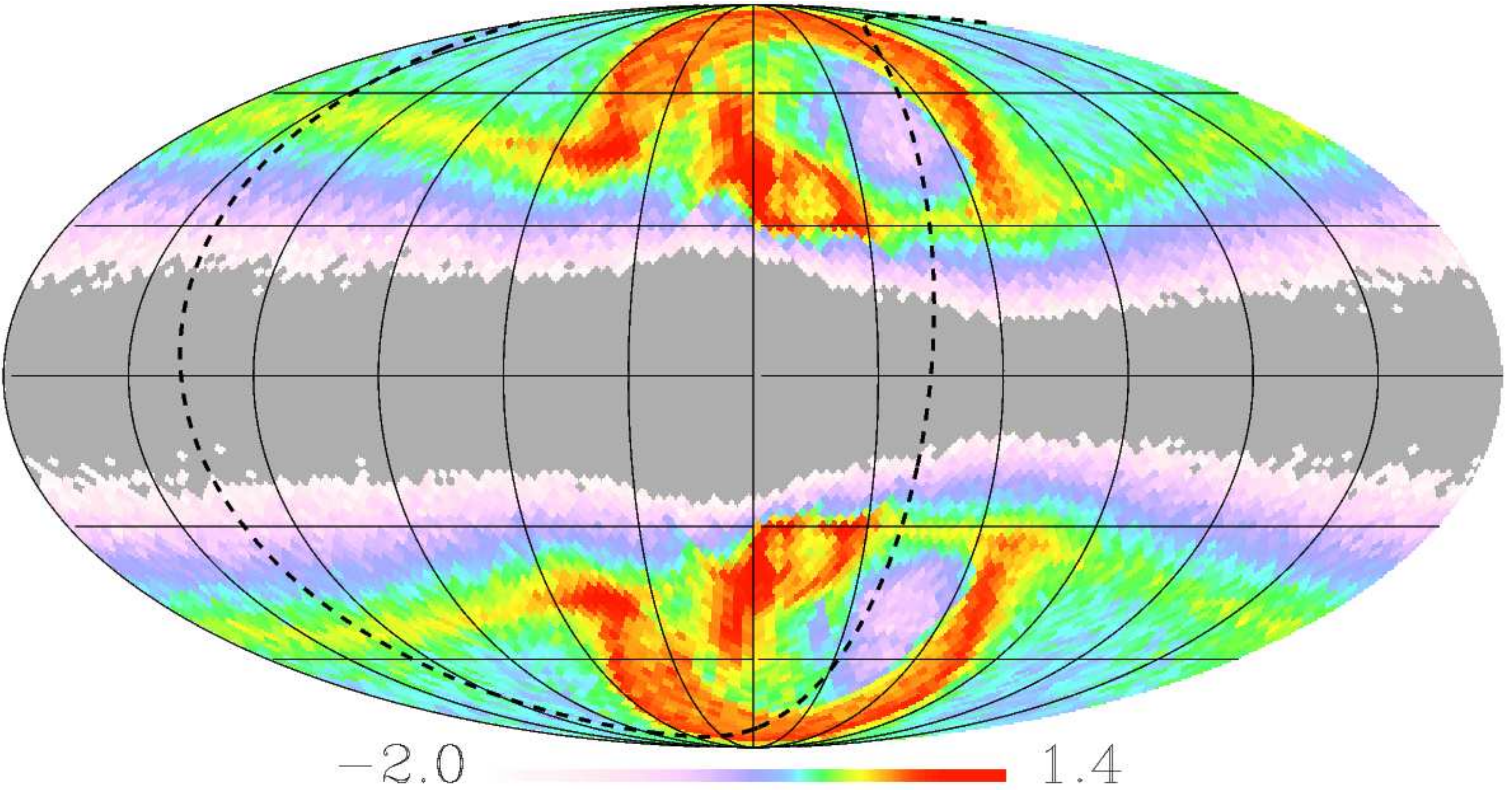}
\includegraphics[width=0.4\textwidth]
{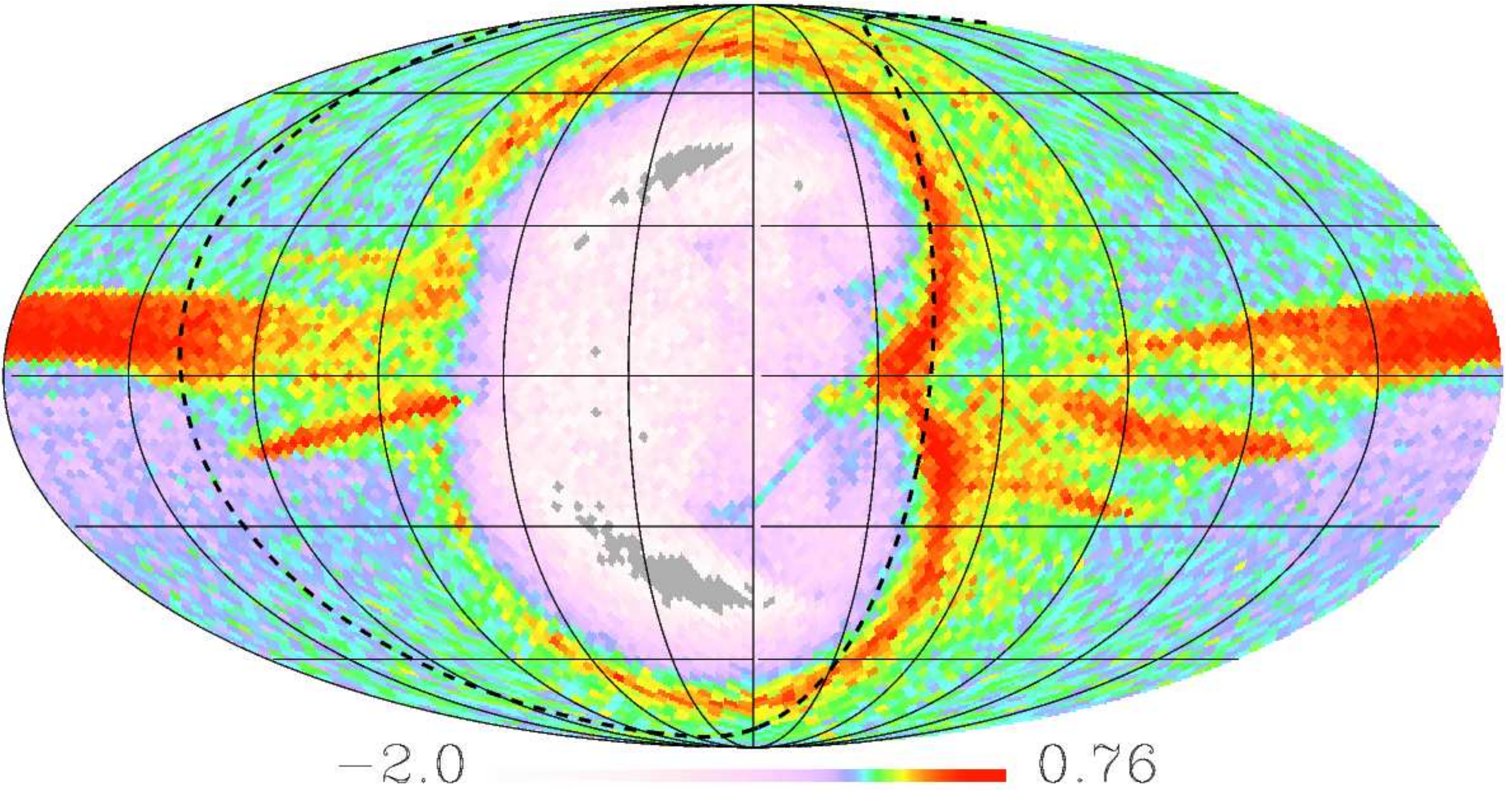}
\caption{Sky maps of the $\Lambda$ values
for Models~I (upper),~II, and~III (lower), 
and \textit{carbon} nuclei. Earth UHECR 
detectors are (almost) blind to the regions in gray. 
The dashed line denotes the SGP.}
\label{f:EGExposureMapCarbon}
\end{figure}

(De-)focusing effects of the Galactic magnetic field  
make the correspondence between the CR arrival directions 
on Earth, and the EG sources contributing to the CR flux 
entering the Galaxy non-trivial~\cite{harari,kachelriess,alvarez}. 
We have constructed maps of parameter $\Lambda \equiv \log_{10}$ of the ratio 
of the number of CR arrivals on the halo border to the one on Earth, using
the HEALPix equal area celestial sphere pixelization~\cite{healpix}.
These maps are shown on Fig.~\ref{f:EGExposureMapCarbon}
for the case of primary \textit{carbon} nuclei. 

Due to these effects, some regions of the
sky have increased probability to contribute to the cosmic rays
observed on Earth, and some others are disfavored.
The non-uniform character of the mapping strengthens in the case
of heavy composition~\cite{vorobiov}, 
so that the regions that would effectively contribute 
to the CR flux on Earth represent only a small fraction 
of the $4\pi$ solid angle.

\subsection{Correlation scan using backtracked directions}
\label{s:ScanParameters}

To quantify the lensing GMF effects on the EG exposure
for the Pierre Auger Observatory, we performed 
a correlation analysis using \textit{backtracked} arrival directions
of simulated events,
and 694 AGN at redshift $z_{\text{max}} \le 0.024$ 
from the VCV catalogue~\cite{vcv}. The simulated CR have been divided sequentially
into samples with the same number of events (81) above
$\unit[40]{EeV}$ as in the Auger data.
The employed set and ranges of parameters were also identical to the ones 
from~\cite{augeragncorrelationSci,augeragncorrelationAPh}.

\begin{figure}[!h]
\centering
\includegraphics[width=0.38\textwidth]
{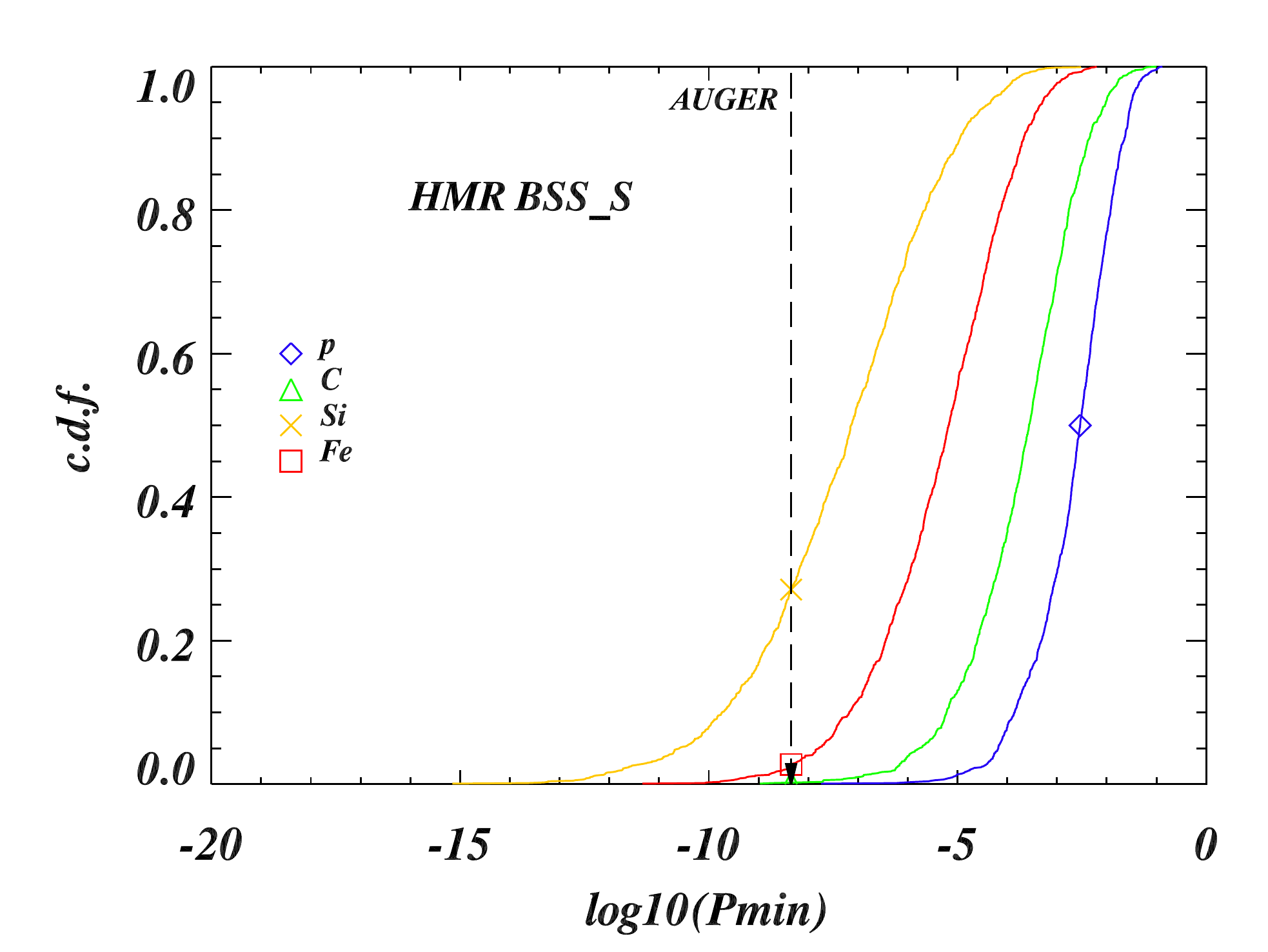}
\caption{Scan results (c.d.f. of $\log_{10}P_{\text{min}}$) of the
simulated CR samples, backtracked under Model~I and 4 masses.
The dashed line shows the value from the Auger 
scan~\cite{augeragncorrelationSci,augeragncorrelationAPh} \textit{on Earth}.
C.d.f. for protons (shown by symbol at the median value) does not intersect this line.}
\label{f:PlotScanResMC_HMRBSSS}
\end{figure}

We will focus here on the minimum probability values, obtained during
the scan. The c.d.f. of decimal logarithm of the corresponding cumulative binomial
probability $P_{{\text{min}}}$ of reaching this level of correlation 
under isotropy for Model I are shown on 
Fig.~\ref{f:PlotScanResMC_HMRBSSS}. The level of the minimum
probability reached in the Auger data is also indicated. 
The results of the correlation scan for the assumed 
LS GMF models and primary mass compositions are summarized in 
Table~\ref{t:DeflectionAndLog10Pmin}.

\begin{table}[!h]
\centering
\begin{tabular}{l||c|c||c|c|c}
\multicolumn{1}{c||}{}&\multicolumn{2}{c||}{\textbf{Deflections}}&\multicolumn{3}{c}{\textbf{Scan results}}
\\
\hline
 & $\vartheta_{50\%}$ & $\vartheta_{95.45\%}$ & $f_{5\%}$ & $f_{50\%}$ & $f_{95\%}$
\\
\hline
\multicolumn{6}{c}{\textbf{HMR bisymmetric even parity model}}
\\
\hline
 p &     4.1 &     7.4 &    -4.2 &    -2.5 &    -1.5\\
 C &    23.7 &    53.4 &    -5.8 &    -3.6 &    -2.0\\
Si &    52.3 &   117.3 &   -10.6 &    -7.1 &    -4.3\\
Fe &    74.5 &   162.4 &    -7.8 &    -5.2 &    -3.3\\
\hline
\multicolumn{6}{c}{\textbf{HMR axisymmetric odd parity model}}
\\
\hline
 p &     4.3 &    10.3 &    -4.5 &    -2.6 &    -1.6\\
 C &    25.5 &    84.1 &    -8.9 &    -5.8 &    -3.7\\
Si &    65.0 &   141.8 &   -12.6 &    -8.7 &    -5.5\\
Fe &    81.1 &   146.3 &    -9.6 &    -6.3 &    -4.0\\
\hline
\multicolumn{6}{c}{\textbf{PS model version by Kachelrie\ss~\etal}}
\\
\hline
 p &     3.0 &    30.1 &    -4.2 &    -2.5 &    -1.5\\
 C &    17.6 &    76.2 &    -4.3 &    -2.5 &    -1.4\\
Si &    37.4 &    92.0 &    -4.9 &    -3.0 &    -1.7\\
Fe &    58.3 &   124.7 &    -4.2 &    -2.6 &    -1.5\\
\end{tabular}
\caption{Magnetic deflections (in $^\circ$), and
$\log_{10}P_{\text{min}}$ values
at the indicated percentiles of the c.d.f.,
for the assumed GMF models 
and primary mass.}
\label{t:DeflectionAndLog10Pmin}
\end{table}

Since for the different GMF model/primary mass assumptions the backtracked directions 
correlate with the catalogue objects in particular privileged regions in the sky, 
our scan results depend strongly on those assumptions. 
For Model~III, the scanned probability
minimum is significantly less deep than for the two other 
spiral-field-only models for any assumed primary mass, 
except for protons, where one obtains nearly the same level of
correlation for all models.  
This model is clearly less compatible with the observed correlation
with the nearby AGN~\cite{augeragncorrelationSci,augeragncorrelationAPh}, 
unless the UHECR flux contribution from these objects (or
objects with similar spatial distribution) is highly non-uniform.

\section{Conclusions and outlook}
\label{s:Conclusions}

Our studies of the UHECR propagation in the LS GMF show
that the UHECR picture observed on Earth is sensitive
to the assumptions on the field structure and/or primary CR composition.

The exact deflection value in the regular component 
of the field depends strongly on the arrival direction on Earth of a cosmic ray,
with the corresponding position angle of deflection differing from one assumed
field distribution to another. To discriminate between GMF models,
additional hints can be provided by the aligned structures of events 
coming from a UHECR source~\cite{multiplets}.

Though the reconstruction of the field is easier in the case of light
primary mass composition, in the case of heavy nuclei the lensing effects 
of the Galactic field on the exposure bring stronger constraints 
on the list of potential UHECR source candidates. The presented analysis 
of the correspondence between the arrival direction distributions on Earth
and at the Galactic border will be complemented by the forward-tracking of cosmic rays 
for a number of plausible UHECR sources scenarios. This will allow for direct
comparison with the observed AGN correlation.


\begin{thebibliography}{99}

\def\vyp#1#2#3{\textbf{#1} (#2) #3}
\def\arxiv#1{[arXiv:#1]}
\def\fullauthor{J.~Abraham \etal}

\bibitem{beck} R.~Beck, 
Space Sci.\ Rev., \vyp{99}{2001}{243}, astro-ph/0012402.

\bibitem{widrow} L.~M.~Widrow, 
RMP\ \vyp{74}{2002}{775}, astro-ph/0207240.

\bibitem{han} J.~L.~Han \etal, 
ApJ\ \vyp{642}{2006}{868}, astro-ph/0601357.

\bibitem{brown} J.~C.~Brown \etal, 
ApJ\ \vyp{663}{2007}{258}, arXiv:0704.0458.

\bibitem{men} H.~Men \etal,
 A\&A\ \vyp{486}{2008}{819}, arXiv:0805.3454.

\bibitem{auger} \verb+http://www.auger.org+ .

\bibitem{augeragncorrelationSci} \fullauthor,
Science \vyp{318}{2007}{939}, arXiv:0711.2256.

\bibitem{augeragncorrelationAPh} \fullauthor,
APh\ \vyp{29}{2008}{188}, arXiv:0712.2843.

\bibitem{augerspectrumPRL} \fullauthor,
PRL 101 (2008) 061101, arXiv:0806.4302.

\bibitem{numerical} W.~H.~Press \etal, 
ISBN 0-521-43108-5.

\bibitem{armengaud} E.~Armengaud, 
PhD, Univ. Paris 7 (2006).

\bibitem{sommers} P.~Sommers, 
Astropart.\ Phys.\ \vyp{14}{2001}{271}, astro-ph/0004016.

\bibitem{AugerSDExposure} 
D.~Allard for the Pierre Auger Collaboration,
Proc.~29$^{\mathrm{th}}$ ICRC, paper no. 134, astro-ph/0511104.

\bibitem{harari} D.~Harari \etal, 
JHEP \vyp{08}{1999}{022}, astro-ph/9906309.

\bibitem{prouza} M.~Prouza and R.~\v{S}m\'ida, 
A\&A\ \vyp{410}{2003}{1}, astro-ph/0307165.

\bibitem{kachelriess} M.~Kachelrie\ss ~\etal,
APh \vyp{26}{2006}{378}, astro-ph/0510444.

\bibitem{vorobiov} S.~Vorobiov \etal,
2009, arXiv:0901.1579.

\bibitem{tinyakovrandom} P.~Tinyakov, I.~Tkachev,  
APh \vyp{24}{2005}{32}, astro-ph/0411669.

\bibitem{alvarez} J.~Alvarez-Mu\~niz \etal,
ApJ\ \vyp{572}{2002}{185}, astro-ph/0112227.

\bibitem{healpix} K.~M.~G\'orski~\etal, 
ApJ\ \vyp{622}{2005}{759}, astro-ph/0409513.

\bibitem{vcv} M.-P.~V\'eron, P.~V\'eron, 
A\&A\ \vyp{455}{2006}{773}.

\bibitem{multiplets} 
D.~Harari~\etal,
JHEP \vyp{0207}{2002}{006}, astro-ph/0205484.
 
\end{thebibliography}
\end{document}